\documentclass[epj]{webofc}
\usepackage[varg]{txfonts}   
\usepackage{graphicx}
\usepackage{amsmath}
\usepackage{comment}

\woctitle{13th International Conference on Meson-Nucleon Physics and the 
Structure of the Nucleon (MENU 2013)}
\begin{document}
\title{Coherent deeply virtual Compton scattering off $^3$He
and
neutron generalized parton distributions}
%
%

\author{Matteo Rinaldi 
\inst{1}\fnsep\thanks{\email{matteo.rinaldi@pg.infn.it}} \and
Sergio Scopetta 
\inst{1}\fnsep\thanks{\email{sergio.scopetta@pg.infn.it}} 
}

\institute{Dipartimento di Fisica e Geologia, Universit\`a degli studi di
Perugia and INFN 
sezione di Perugia, Via A. Pascoli 06100
Perugia, Italy
}

\abstract{%

It has been recently proposed to study coherent deeply virtual
Compton scattering (DVCS) off
$^3$He nuclei to access neutron generalized parton 
distributions (GPDs). In particular,
it has been shown that, in Impulse Approximation (IA) 
and at low momentum transfer, the sum of the quark helicity conserving
GPDs of $^3$He, $H$ and $E$, is dominated by the neutron contribution. This 
peculiar result makes the $^3$He target very promising to access the neutron 
information. We present here
the IA calculation of  the spin dependent
GPD $\tilde H$ of $^3$He.  Also for this quantity the neutron 
contribution is found to be the dominant one, at low momentum transfer.
The known forward limit of the IA calculation of $\tilde H$,
yielding the polarized parton distributions of $^3$He, is correctly
recovered.
The extraction of the neutron information could be anyway 
non trivial, 
so that a procedure, able to take into account the nuclear 
effects encoded in the IA analysis, is proposed.
These calculations, essential for the evaluation of the coherent DVCS 
cross section asymmetries, which depend on the GPDs $H, E$ and $\tilde H$,
represent a crucial step for planning possible experiments at Jefferson Lab.
}
\maketitle

Generalized parton distributions (GPDs), introduced in Ref. \cite{uno},
describe the hadron non perturbative structure in
hard exclusive processes. 
The study of these quantities
is very important to understand several open issues of hadrons,
such as, $e.g.$, their spin structure.
GPDs are
related to the total angular momentum of partons inside the hadron;
their knowledge, once
the helicity quark contribution were subtracted, 
would make possible to access
their orbital angular momentum (OAM) content.

\begin{figure*}[t]
\vspace{6.5cm}
\includegraphics{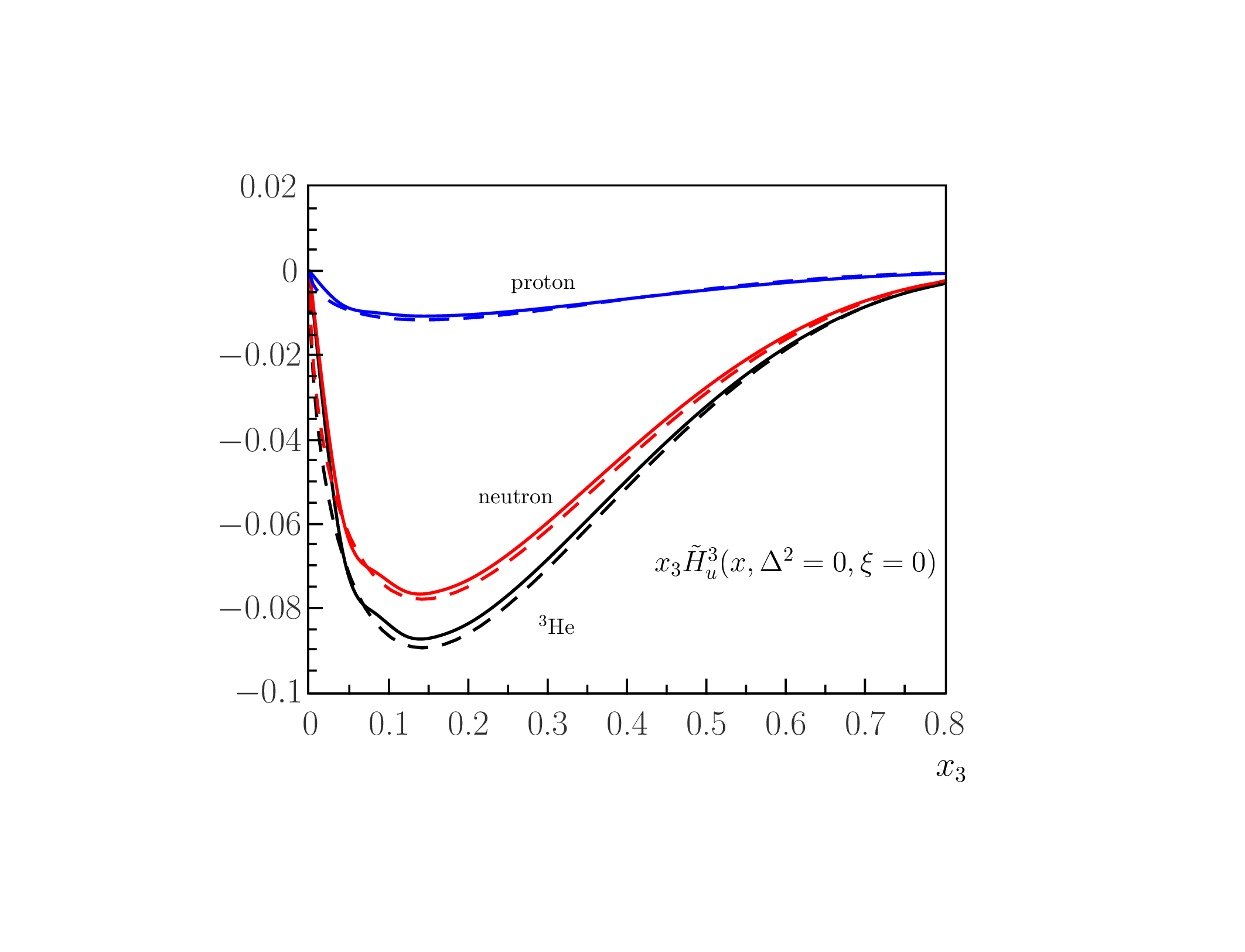}
\includegraphics{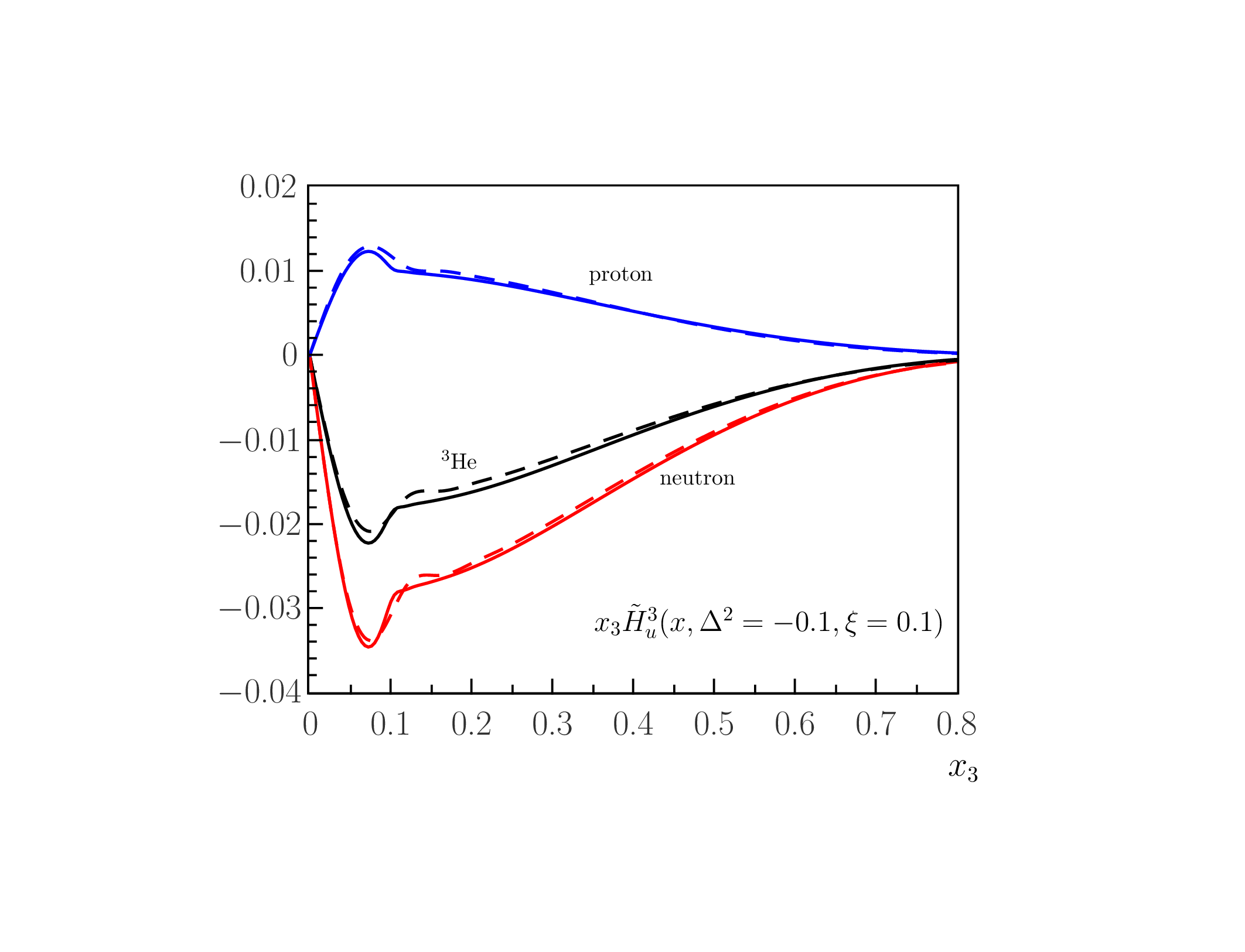}
%
\caption{ 
The  GPD $x_3 \tilde{H}^{3,u}(x,\Delta^2,\xi)$ for the flavor $q=u$, 
where $x_3 = (M_3/M) x$
and $\xi_3 = (M_3/M) \xi$, shown in the forward limit (left panel)
and at $\Delta^2 = -0.1
\ \mbox{GeV}^2$ and $\xi_3=0.1$ (right panel), 
together with the neutron and the proton
contribution. Solid lines represent the full IA result, Eq. (\ref{new}),
while the dashed ones correspond to the approximation
Eq. (\ref{sgu}).}
\label{fig-1} 
\end{figure*}

The most promising process for these studies is likely
deeply virtual compton
scattering (DVCS),
$i.e.$, the reaction
$eH \longmapsto e'H' \gamma$ \ when \ $Q^2\gg M^2$ ($Q^2=-q \cdot q$ \ 
is the momentum transfer
between the leptons $e$ and 
$e'$, $\Delta^2$ the one between hadrons $H$ and $H'$ with
momenta $P$ and $P'$, and
$M$ the nucleon mass).
Another relevant kinematical variable is the so called
skewedness, $\xi = - \Delta^+/(P^+ + P^{'+})$ 
\footnote{In this paper, $a^{\pm}=(a^0 \pm a^3)/\sqrt{2}$.}.

The measurement of GPDs is a hard experimental challenge; despite of this,
data for proton and nuclear targets are being analyzed, see, $e.g.$, Refs.
\cite{data1,data2}.

In particular, the study of nuclear GPDs is a relevant task to obtain new
information on possible modifications of the bound nucleon structure (this
analysis started in
Ref. \cite{deu}), and also to extract neutron GPDs, 
necessary to get, together with the proton ones, a flavor
decomposition of GPDs. For this kind of studies,  $^3$He   is a very
important target due to its spin structure (see, $e.g.$, Refs.
\cite{3He,old}).
In fact, among the light nuclei, $^3$He is the
only target for which the sum $\tilde{G}_M^{3,q}(x,\Delta^2,\xi) =
H^{3}_q(x,\Delta^2,\xi)+
E^{3}_q(x,\Delta^2,\xi)$ could be dominated by the neutron contribution.
In facts,
$^2$H and $^4$He are isoscalar and not useful to this
aim,  see Ref. \cite{noi}. In this latter work it has been also shown 
to what extent this fact can be used to extract the neutron information.

The details of the calculation of $^3$He GPDs 
in IA can be found in Ref. \cite{scopetta},
where the GPD $H$ of $^3$He,
$H_q^3$, has been obtained
as a convolution-like equation
in terms of the corresponding
nucleon quantity.
The treatment has been later extended to
$\tilde{G}_M^{3,q}$
(see Ref. \cite{noi} for details),
and to $\tilde{H}^3_q$,
giving the following result:
\begin{eqnarray}
 ~~~~~~ \tilde G_M^{3,q}(x,\Delta^2,\xi)  = 
\sum_N
\int dE 
\int d\vec{p}~
{\left [ P^N_{+-,+-}
-
P^N_{+-,-+} \right](\vec p,\vec p\,',E) }
{\xi' \over \xi}
\tilde G_M^{N,q}(x',\Delta^2,\xi'),
\end{eqnarray}
\vskip-3mm
\noindent
and
\begin{eqnarray}
 ~~~~~~{\tilde H_{q}^3(x,\Delta^2,\xi)}  = 
\sum_N
\int dE 
\int d\vec{p}
\,
{\left [ P^N_{++,++}
-
P^N_{++,--} \right](\vec p,\vec p\,',E) }
{\xi' \over \xi}
{\tilde H^{N}_q(x',\Delta^2,\xi')}~,
\label{new}
\end{eqnarray}
\vskip-3mm
\noindent
respectively.

Here, $x'$ and $\xi'$ are the variables 
for the bound nucleon 
GPDs, where $p \, (p'= p + \Delta)$
is its 4-momentum in the initial (final) state and, eventually, 
proper components appear of the spin dependent,
one body off diagonal spectral function:

\begin{eqnarray}
 \label{spectral1}
 ~~~~~~ P^N_{SS',ss'}(\vec{p},\vec{p}\,',E) 
= 
\dfrac{1}{(2 \pi)^6} 
\dfrac{M\sqrt{ME}}{2} 
\int d\Omega _t
\sum_{\substack{s_t}} \langle\vec{P'}S' | 
\vec{p}\,' s',\vec{t}s_t\rangle_N
\langle \vec{p}s,\vec{t}s_t|\vec{P}S\rangle_N~,
\end{eqnarray}
where $S,S'(s,s')$ are the nuclear (nucleon) spin projections
in the initial (final) state, respectively,
and $E= E_{min} +E_R^*$, 
being $E^*_R$ the excitation energy 
of the full interacting two-body recoiling system.
The main quantity appearing in the definition
Eq. (\ref{spectral1}) is
the intrinsic overlap integral
\begin{equation}
\langle \vec{p} ~s,\vec{t} ~s_t|\vec{P}S\rangle_N
=
\int d \vec{y} \, e^{i \vec{p} \cdot \vec{y}}
\langle \chi^{s}_N,
\Psi_t^{s_t}(\vec{x}) | \Psi_3^S(\vec{x}, \vec{y})
 \rangle~
\label{trueover}
\end{equation}
between the wave function
of $^3$He,
$\Psi_3^S$,  
and the final state, described by two wave functions: 
{\sl i)}
$\Psi_t^{s_t}$, 
eigenfunction 
with eigenvalue
$E = E_{min}+E_R^*$, of the state with intrinsic
quantum numbers $s_t$ of the
Hamiltonian pertaining to the system of two {\sl interacting}
nucleons with relative momentum $\vec{t}$, 
which can be either
a bound 
or a scattering state, and 
{\sl ii)}
the plane wave representing 
the nucleon $N$ in IA.
A numerical evaluation of Eqs. (1) and (2),
requires the knowledge of
the overlaps, Eq. (4), appearing in Eq. (3).
Use has been made of those
corresponding to the analysis presented in Ref.
\cite{overlap} in terms of 
AV18  \cite{pot} wave functions
\cite{AV18}.

In order to numerically evaluate Eqs. (1-2), A model
for the free nucleonic GPDs has to be chosen. For $H$ and $E$ 
use was made of a simple model, that of
Ref. \cite{Rad1}), later minimally extended to
evaluate and describe the spin dependent GPD $\tilde H$ (see
Ref. \cite{noi} for details). In order to confirm the validity of the
calculation, since  data for $^3$He GPDs are not available, it is necessary
to verify the usual limits of the GPDs, $i.e.$, the forward limit and the
first
moment.

In Ref. \cite{scopetta} it was shown that the calculation 
of $H^3_q(x, \Delta^2,\xi)$ fulfills these constraints , 
while for  $\tilde G^3_q(x, \Delta^2,\xi)$, 
since the forward limit of the GPD $E$ is not defined, the only
available check is the first moment. In particular, we have calculated 
$
\sum_q \int dx \, \tilde G_M^{3,q}(x,\Delta^2,\xi) = G_M^3(\Delta^2)
$, which is the well known magnetic form factor (ff) of $^3$He.
Our result is found to be in perfect agreement  with previous
calculations ($e.g$. the one-body part of the
AV18 calculation
presented in Ref. \cite{schiavilla}) and, for 
$-\Delta^2 \le 0.15$ GeV$^2$, which is the relevant kinematic region for
the coherent DVCS at JLab, our results compare well also with the data.
In the $\tilde H^3_q$ case, we found that our calculation, in the forward
limit, reproduces formally and numerically the results of Ref.
\cite{old} for polarized DIS off $^3$He. Anyway it is necessary to remark
that the first moment of $\tilde H^3_q$ is related to the axial ff off
$^3$He, which is poorly known, so that this check is not really useful for
comparison. After this positive check,
we analyze next the proton and 
neutron contribution to $^3$He GPDs. In particular,  $\tilde H^3_q$,
related to a polarized target ,
should be dominated by the neutron contribution.
Let us show now to what extent
this feature is obtained and how, thanks to this observation,
the neutron information can be extracted.
In Fig.\ref{fig-1} are shown numerical results of Eq.(2) for two different
values of momentum transfer. The neutron contribution is found to largely
dominate the $^3$He GPD, but increasing $\Delta^2$ the proton contribution
grows up (see Fig.\ref{fig-1}, solid lines in both panels), in particular
for $u$ flavor. Therefore, it is necessary  to introduce
a procedure to safely extract the neutron information from 
$^3$He
data. This can be obtained by noticing that that Eq.(2) can be written:

\begin{eqnarray}
 ~~~~~~ ~~~~~~ ~~~~~~ ~~~~~~\tilde H^{3}_q(x,\Delta^2,\xi) =   
\sum_N \int_{x_3}^{M_A \over M} { dz \over z}
h_N^3(z, \Delta^2, \xi ) 
\tilde H^{N}_q \left( {x \over z},
\Delta^2,
{\xi \over z},
\right)~,
\end{eqnarray}

where $h_N^3(z, \Delta^2, \xi )$ 
is a ``light cone spin dependent off-forward momentum
distribution'' which
is strongly peaked around $z=1$
close to the forward limit. 
Therefore, in this region,
for $x_3 = (M_A/M) x \leq 0.7$ one has:

\begin{eqnarray}
 ~~~~~~ ~~~~~~ ~~~~~~ ~~~~~~ \tilde H^{3}_q(x,\Delta^2,\xi) 
& \simeq &   
\sum_N 
\tilde H^{N}_q \left( x, \Delta^2, {\xi } \right)
\int_0^{M_A \over M} { dz }
h_N^3(z, \Delta^2, \xi ) 
\nonumber
\\
& = &
G^{3,p,point}_A(\Delta^2) 
\tilde H^{p}_q(x, \Delta^2,\xi) 
+ 
G^{3,n,point}_A(\Delta^2) 
\tilde H^{n}_q(x,\Delta^2,\xi)~. 
\label{sgu}
\end{eqnarray}
\vskip-1mm
Here, the axial point like ff, 
$G^{3,N,point}_A(\Delta^2)=\int_0^{M_A \over M} dz \, h_N^3(z,\Delta^2,\xi),
$ 
which would give the nuclear axial ff if the proton and the neutron were 
point-like systems, have been introduced.

For small values of $\Delta^2$ these quantities depend slowly on the
nuclear potential , so that they are under good theoretical control. It is
important to remark that, in the forward limit, they reproduce the so called 
``effective polarizations''  of the protons ($p_p$) 
and the neutron ($p_n$) in $^3$He,
whose values are rather similar 
if evaluated within different nucleon nucleon potentials 
(see Refs. \cite{3He,old,overlap} for details). For the AV18 potential,
used in this work, 
the values $p_n=0.878$ and $p_p= -0.023$ are obtained. Now we can arrange 
Eq.(6) to extract  the neutron GPD from possible future data for proton and
$^3$He:

\vskip-5mm
\begin{eqnarray}
\label{extr}
 ~~~~~~ ~~~~~~\tilde H^{n,extr}(x, \Delta^2,\xi)  \simeq  
{1 \over G^{3,n,point}_A(\Delta^2)} 
\left\{ \tilde H^3(x, \Delta^2,\xi) 
 -  
G^{3,p,point}_{A}(\Delta^2) 
\tilde H^p(x, \Delta^2,\xi) \right\}~.
\end{eqnarray}

We have compared the free neutron GPD of the used model  with the
numerical evaluation of Eq.(7), using the same nucleonic model for $\tilde
H_p$ and the full calculation of $\tilde H^3$.  Fig.\ref{fig-1} (right
panel) shows that the procedure  works, as we expected, in particular  for 
$x \leq 0.7$, which is the relevant region for DVCS at JLab. It is
remarkable that the only theoretical ingredients of the calculation
are the axial point like ffs. Our results confirm that coherent DVCS off
$^3$He, at low momentum transfer $\Delta^2$, is an ideal process to obtain 
the neutron GPDs.
If the experimental
kinematic region were extended at higher $\Delta^2$, a relativistic treatment 
\cite{ema}
and/or the inclusion of many body currents, beyond the present IA scheme, 
should be implemented.
Now we have at hand all the ingredients for the evaluation of the 
cross section
asymmetries, at leading twist, for a  spin 1/2 target.
This analysis is going on using the formalism introduced in Ref.\cite{diet}.

This work was supported in part by the Research Infrastructure
Integrating Activity Study of Strongly Interacting Matter (acronym
HadronPhysic3, Grant Agreement n. 283286) under the Seventh Framework
Programme of the European Community.


\begin{thebibliography}{}
%
%

\bibitem{uno}
Mueller,~D. {\sl et al.},
Fortsch.\ Phys.\  {\bf 42}, 101 (1994)
\bibitem{due}
Radyushkin,~A.~V.,
Phys.\ Lett.\ B {\bf 380}, 417 (1996);
Ji,X.~-D.,
Phys.\ Rev.\ Lett.\  {\bf 78}, 610 (1997)
  

\bibitem{data1}
Mazouz,~M. {\it et al.} , [Jefferson Lab Hall A Collaboration],
Phys.\ Rev.\ Lett.\  {\bf 99} 242501 (2007)

\bibitem{data2}
Guidal,~M,
  Phys.\ Lett.
\ B {\bf 693}, 17 (2010)
  


\bibitem{deu} 
  Berger,~E.~R., Cano,~F., Diehl,~M., and Pire,~B,
  Phys.\ Rev.\ Lett.\  {\bf 87}, 142302 (2001)
  
\bibitem{3He}
Friar,~J.~L. {\it et al.},
  Phys.\ Rev.\ C {\bf 42} 2310 (1990)

\bibitem{old}
Ciofi degli Atti,~C. {\it et al.},
Phys.\ Rev.\ C {\bf 48} 968 (1993)
  
  
\bibitem{noi}
Rinaldi,~M. and Scopetta,~S.,
Phys. Rev. C 85, 062201(R) (2012)
%
Rinaldi,~M. and Scopetta,~S.,

\bibitem{scopetta}
Scopetta,~S.,
  Phys.\ Rev.\ C {\bf 70} 015205 (2004);
Scopetta,~S.,   
Phys.\ Rev.\ C {\bf 79} 025207 (2009)

  
\bibitem{overlap}
Kievsky,~A., Pace,~E., Salm\`e,~G. and Viviani,~M.,
  Phys.\ Rev.\ C {\bf 56}, 64 (1997)

\bibitem{pot} 
Wiringa,~R.~B., Stoks~V.~G.~J. and Schiavilla,~R.,
Phys.\ Rev.\ C {\bf 51} 38 (1995)
  
\bibitem{AV18}
Kievsky,~A., Viviani,~M. and Rosati,~S.,
Nucl.\ Phys.\ A {\bf 577} 511 (1994)
  
\bibitem{Rad1}
Musatov,~I.~V. and Radyushkin,~A.~V.,
Phys.\ Rev.\ D {\bf 61} 074027 (2000)
  
\bibitem{schiavilla}
Marcucci,~L.~E., Riska,~D.~O. and Schiavilla,~R.,
Phys.\ Rev.\ C {\bf 58}, 3069 (1998)
  
\bibitem{ema} 
Pace,~E, Salm\`e,~G., Scopetta,~S., Del Dotto,~A. and Rinaldi,~M.,
Few Body Syst.\  {\bf 54}, 1079 (2013)

\bibitem{diet} 
  Belitsky, A.~V., Mueller, D., Niedermeier, L. and Sch\"afer, A.,
Nucl.\ Phys.\ B {\bf 593}, 289 (2001)





\end{thebibliography}
\end{document}